\documentclass[twocolumn,showpacs,preprintnumbers]{revtex4}
\usepackage{graphicx}
\usepackage{dcolumn}
\usepackage{bm}
 
\begin{document}
 
\preprint{}
\title{Sound Propagation in Nematic Fermi Liquid}
\author{Hae-Young Kee}
\affiliation{Department of Physics, University of Toronto, Toronto,
Ontario M5S 1A7, Canada}
\date{\today}
 
\begin{abstract}
We study the longitudinal sound propagation in the electronic nematic
Fermi liquid where the Fermi surface is distorted due
to the spontaneously broken rotational symmetry.
The behavior of the sound wave in the nematic ordered state is
dramatically different from that in the isotropic Fermi liquid.
The collective modes associated with the fluctuations 
of the Fermi surface distortion in the nematic Fermi liquid
leads to the strong and anisotropic damping of the sound wave.
The relevance of the nematic Fermi liquid in doped Mott insulator 
is discussed.
\end{abstract}
 
\maketitle
 
\def\be{\begin{equation}}
\def\ee{\end{equation}}
\def\bea{\begin{eqnarray}}
\def\eea{\end{eqnarray}}
 
\paragraph{Introduction}
There are compelling experimental evidences that
inhomogeneous or anisotropic collective phases  
exist in strongly correlated electron systems.
The charge ordered phase - called stripes -  
was  directly observed via the electron diffraction images
in transition metal oxides such as manganite, La$_{1-x}$Ca$_x$MnO$_3$.\cite{cheong}
The neutron scattering experiment\cite{tranquada} supports the existence
of the stripe phase in high temperature superconductors.\cite{carlson}
The transport measurements on the ultra-clean two dimensional
electron systems with high magnetic field
reveals the anisotropic metallic state, which
suggests the existence of a novel anisotropic collective phase.\cite{lilly}
The theoretical studies have focused on a unidirectional 
charge or spin density waves
with broken translational symmetry - called smectic phase - 
which can be viewed as the coupled one-dimensional Luttinger liquid.\cite{fradkin}
A nematic state with broken rotational
symmetry  has been considered as the melted smectic phase
due to the topological defects such as dislocations.
While this is a natural route to achieve the nematic state\cite{emery},
we are still lack of the mathematical formalism to describe
the quantum phase transition from the electronic smectic to  
nematic state.

Recently, a different route to the nematic state was proposed in \cite{vadim}.
A microscopic theory of the electronic nematic phase proximate to
an isotropic Fermi liquid was discussed from the weakly correlated side.
It was shown that the quadrupolar density-density
interaction with a negative coefficient leads to the nematic Fermi liquid 
which breaks the rotational symmetry.
As a consequence of the broken rotational symmetry,
the Fermi surface is elongated like an ellipse.
We call the major axis of the elliptical Fermi surface
the principal axis.
The fluctuation of the principal axis of the Fermi surface
is associated with the collective mode.
The Goldstone mode which is the transverse fluctuation of the principal axis
is overdamped in the ordered phase due to the Landau damping.
It was shown that the coupling of quasiparticles to the Goldstone mode
leads to the breakdown of Fermi liquid theory.\cite{vadim}
The pairing instability in the nematic  Fermi liquid was 
also studied in \cite{kim}.
It was shown that the Goldstone mode
mediates the pairing between electrons with 
different spins(or pseudospins), 
despite the repulsive quadrupolar density-density
interaction between electrons with different spins.
The structure of the gap in the paired nematic state
was identified with four nodes on the Fermi surface.
The gap vanishes exponentially near the nodes
which suggests the nontrivial effect on the thermodynamic
quantities in a nematic superconductor.\cite{kim}

In this paper, we study the longitudinal sound propagation
in the nematic Fermi liquid where the Fermi surface is
distorted due to the spontaneously broken rotational symmetry.
In the absence of the coupling to the collective modes, 
the sound velocity shift is reduced 
in the ordered state compared to that in the isotropic Fermi liquid.
Its reduction is anisotropic, and it is maximum along the symmetric directions,
the major and minor axes of the elliptical Fermi surface.
The collective  modes couple to the stress tensor,
which affects the dynamics of the sound propagation.
Since the Goldstone mode is overdamped in the nematic state,
its coupling to the sound wave results in the strong damping
of the sound except along the symmetric directions.
The coupling to the amplitude mode leads to
the additional sound attenuation. 

The paper is organized as follows.
We  review the general formula of the auto-correlation functions
to compute the sound velocity shift and the sound attenuation in section 2.
We show  how sound propagates in the nematic state without taking 
into account the coupling to the collective modes in section 3,
which is presented to compare with the main results shown in section 4. 
In section 4,
the coupling between the sound wave and the collective modes is considered, which
leads to the strong damping of the sound wave.
We summarize the results and discuss the relevance
of the nematic state to a doped Mott insulator in
the context of change of sound attenuation in  section 5. 

\paragraph{Auto-correlation function in collisionless regime}
In the collisionless regime, the sound can propagate
due to the molecular field caused by the surrounding medium 
which produces the stabilizing force.
In solid, the lattice vibration due to externally applied
sound waves induces a scalar potential, which
acts as an external perturbation to the electron system.
The relation between the stress tensor and the dynamics of
the sound waves is extensively discussed in the work of
Kadanoff and Falko.\cite{kadanoff}
Thus we present a brief summary of the auto-correlation function.

For the longitudinal sound wave,
the sound velocity shift, $\delta C_l$, 
and the attenuation, $\alpha_l$, at low frequencies can be
computed from the auto-correlation function, $\langle [h_l,h_l]\rangle$
as follows.
\begin{equation}
{\delta C_l \over C_l} 
= - \left. {\omega  \over  m_{\rm ion} C_l |{\bf q}|}
{\rm Re} \langle [h_l,h_l] \rangle
({\bf q},\omega) \right |_{\omega=C_l |{\bf q}|} \ ,
\end{equation}

\begin{equation}
\alpha_l = \left. {\omega \over m_{\rm ion} C_l |{\bf q}|}
{\rm Im} \langle [h_l,h_l] \rangle
({\bf q},\omega) \right |_{\omega=C_l |{\bf q}|} \ ,
\end{equation}
where
\begin{equation}
h_l({\bf r},t) = \frac{q}{\omega} \tau_{xx}({\bf r},t)-
\frac{\omega m}{q} n({\bf r},t).
\end{equation}
$C_l$ is the sound velocity, and
$m_{\rm ion}$ and $m$ are the mass of ions and the mass of electron,
respectively.
$n ({\bf r},t)$ is the density operator, and $\tau_{ij}$ is
the stress tensor operator. 
The auto-correlation function, $\langle [\tau_{ij}, \tau_{ij}] \rangle$,
of the stress tensor $\tau_{ij}$ is defined as
\begin{equation}
\langle [\tau_{ij}, \tau_{ij}] \rangle ({\bf r}-{\bf r'}, t-t')
\equiv - i \theta (t-t')
\langle [\tau_{ij} ({\bf r},t), \tau_{ij} ({\bf r'},t') ]
\rangle \ .
\end{equation}
Here the stress tensor, $\tau_{ij}$ is written as 
\begin{equation}
\tau_{ij} ({\bf r},t) = \sum_{\sigma}
\left [
{(\nabla-\nabla^{'})_i \over 2i}{(\nabla-\nabla^{'})_j \over
2im} \psi^{\dagger}_{\sigma} ({\bf r},t)
\psi_{\sigma} ({\bf r'},t)
\right ]_{{\bf r'}={\bf r}} \ ,
\end{equation}
where $\psi^{\dagger}_{\sigma}$ is the electron creation
operator with spin $\sigma$.


\paragraph{Sound propagation in a nematic Fermi liquid}
We consider the following Hamiltonian for a two dimensional fermion
system.
\begin{eqnarray}
H &=& \int d^2 r 
 \psi^{\dagger}_{\sigma} ({\bf r}) 
\epsilon(\nabla) \psi_{\sigma}({\bf r}) \cr
&+& \int d^2 r d^2 r'  F_2 ({\bf r}-{\bf r'})
{\rm Tr} [{\hat Q}_{\sigma} ({\bf r})  
{\hat Q}_{\sigma} ({\bf r'}) ]\,
\end{eqnarray}
where $\epsilon (\nabla)$ is the kinetic energy operator.
The matrix quadrupolar density of the fermions is defined as
\begin{equation}
{\hat Q}_{\sigma} ({\bf r}) = - {1 \over k_F^2}
\psi^{\dagger}_{\sigma} ({\bf r})
\left (\matrix{\partial^2_x - \partial^2_y & 2 \partial_x \partial_y \cr
2 \partial_x \partial_y & \partial^2_y - \partial^2_x \cr} \right )
\psi_{\sigma} ({\bf r}).
\end{equation}
The interactions, $F_2 ({\bf q})$ between the fermions, is given by
\begin{equation}
F_2 ({\bf q}) = \frac{F_2}{1+\kappa q^2}.
\end{equation}
The details of the form of the interaction do not matter 
in describing the occurrence of the nematic Fermi liquid, 
as far as the short range interaction is concerned. 

Under the condition $F_2^{-1} +N(0)/2 < 0 $, the nematic order occurs.
Using the mean-field approximation,
the band dispersion in the nematic state is given by
\begin{equation}
\xi_{\bf k}= {\bf v_F}\cdot {\bf k} (1+ a (k/k_F)^2) 
+  {\cal N} \cos{(2\theta_k)},
\end{equation}
where ${\bf k}$ is measured from the Fermi surface,
and ${\cal N}$ is is determined from the expectation value of 
the order parameter $\langle Q_{xx} \rangle$ at the saddle point 
as follows.\cite{vadim2}
\begin{eqnarray}
\langle Q_{xx} \rangle
& = &\sum_{\bf k} \cos{(2\theta_k)} f(\xi_{\bf k})
= {\cal N} N(0) \left( \frac{1}{2}- \frac{3 a {\cal N}^2}{8 \epsilon_F^2}
\right),
\nonumber\\
\langle Q_{xy} \rangle  &= &0.
\end{eqnarray}
where $\epsilon_F = v_F k_F$,  and $N(0)$ is the density of state.
Note that the principal axis of the distorted Fermi surface is set as ${\hat x}$.

Using the single particle Green function in the nematic state given by,
\begin{equation}
G^{-1} (\omega, {\bf k}) = i\omega_n -\xi_{\bf k},
\end{equation}
the auto-correlation function of the stress tensor can be obtained
as follows.
\begin{eqnarray}
\langle [\tau_{ll},\tau_{ll}] \rangle (\omega, {\bf q})
&=& \frac{1}{\beta} \sum_{i \nu_n}
\sum_{\bf k} \frac{p_F^4}{m^2} \cos^4{(\theta-\phi)}
\nonumber\\
& & \hspace{-0.5cm} \times
G(i\nu_n,{\bf k}) G(i\omega_n+i \nu_n, {\bf k}+{\bf q})
\nonumber\\
&& \hspace{-2.8cm} = \frac{N(0) p_F^4}{4 m^2} \left(
\frac{3}{2}- \frac{15 a {\cal N}^2}{8 \epsilon_F^2} 
- \frac{3 a  {\cal N}^2}{4 \epsilon_F^2} \cos^2{(2\phi)}
-2 i s 
\right),
\end{eqnarray}
where $\theta$ and $\phi$ are the angles
of ${\bf k}$ and ${\bf q}$ from the ${\hat x}$-axis, respectively,
and $s= \omega/(v_F |{\bf q}|)$.
%
%
The density-density correlation function, $\langle [n,n]\rangle$
and the cross term $\langle [\tau_{ll}, n] \rangle$,
are smaller than the correlation function of stress tensor
by $(C_l/v_F)^2 $ and $(C_l/v_F)^4$, respectively.
Notice that the ratio of the sound velocity ($C_l$) and the Fermi velocity
($v_F$), $C_l/v_F << 1$ in a metal.  
Therefore, the sound velocity change is given by
\begin{equation}
\frac{\delta C}{C_l} = -\lambda_l 
\left( \frac{3}{2}- \frac{15 a {\cal N}^2}{8 \epsilon_F^2} 
- \frac{3 a {\cal N}^2}{4 \epsilon_F^2} \cos^2{(2\phi)} \right),
\end{equation}
where $\lambda_l = N(0) p_F^4/(4 m^2 m_{ion})$.
The above result in the limit, ${\cal N} \rightarrow 0$, recovers the 
sound velocity shift in the isotropic Fermi liquid 
when the long-range Coulomb interaction is not taken into account.\cite{note}

We find that without taking into account the coupling to the collective modes,
the sound wave is weakly damped just like the situation
in the isotropic Fermi liquid state which is order of
$\lambda_l (C_l/v_F)$. 
However, the anisotropic reduction of the sound velocity shift occurs 
in the nematic ordered state.
The reduction of the sound velocity shift is maximum 
when the wave vector ${\bf q}$ is along the major and minor 
axes of the Fermi surface --- the angle, $\phi$ between the wave
vector ${\bf q}$ and the principal axis of the distorted
Fermi surface, ${\hat x}$, is $0$ along the major axis,
and $\pi/2$ along the minor axis.

\paragraph{Coupling to the collective modes}
In the nematic ordered state, there are two collective modes
associated with the fluctuations of the distorted Fermi surface.
Since the collective modes couple to the sound wave through
the stress tensor, the auto-correlation function is renormalized due
to the coupling to the collective modes, which can be written as
\begin{equation}
\langle [h_l,h_l]\rangle
=\langle [h_l,h_l]\rangle_{0} +
\sum_{i={\perp},{\parallel}}
\frac{F_2 \langle [h_l,\delta {\cal N}_i]\rangle
\langle [\delta {\cal N}_i,h_l]\rangle}
{ 1-F_2 \langle [\delta {\cal N}_i,\delta {\cal N}_i]\rangle},
\end{equation}
where $\delta {\cal N}_{\perp}$ and $\delta {\cal N}_{\parallel}$
are  the transverse and longitudinal fluctuations of
the distorted Fermi surface, which corresponds to
the Goldstone and amplitude modes, respectively.

\subparagraph{coupling to the Goldstone mode}
The stress tensor directly couples to
the Goldstone mode which is the transverse 
fluctuation of the principal axis of the Fermi surface in the nematic ordered
state. 
The Goldstone mode does not couple to the uniform density
in the static regime, because the nematic order preserves the local volume.
--- there is no change of the volume of the Fermi surface in
the nematic ordered state.

The correlation function of the 
stress tensor and the Goldstone mode, $\delta {\cal N}_{\perp}$, is given by
\begin{eqnarray}
\langle [\tau_{ll}, \delta {\cal N}_{\perp}] \rangle
&= & \frac{N(0) p_F^2}{m} \int \frac{d\theta}{2\pi}
\cos^2{(\theta -\phi)} \sin{(2\theta)}
\nonumber\\
& & \hspace{-1.0cm} \times
\left[ 1 - \frac{ 3 a {\cal N}^2}{\epsilon_F^2} \cos^2{(2\theta)}
-\frac{s}{s-\cos{(\theta-\phi)}} \right]
\nonumber\\
&& \hspace{-1.8cm} = 
\frac{\sin{(2\phi)} N(0) p_F^2}{2m}
\left[ \frac{1}{2}-\frac{3 a {\cal N}^2}{8 \epsilon_F^2} -2 i s
   +O(s^2) \right].
\end{eqnarray}
On the other hand, the Goldstone mode propagator is obtained as\cite{vadim} 
\begin{equation}
\langle [\delta {\cal N}_{\perp}, \delta {\cal N}_{\perp}] \rangle
= 
F_2^{-1} - is  N(0)  \sin^2{(2\phi)}
+O(s^2).
\end{equation}
It is important to notice that the Goldstone mode 
is overdamped due to the anisotropic Landau damping.
Therefore, the coupling to the Goldstone mode
can be written  as 
\begin{eqnarray}
\frac{F_2(q) \langle [h_l,\delta {\cal N}_{\perp}]\rangle
\langle [\delta {\cal N}_{\perp},h_l]\rangle}{ 1-F_2(q) 
\langle [\delta {\cal N}_{\perp},\delta {\cal N}_{\perp} ]\rangle}
&=& \frac{F_2 N(0)^2 \sin^2{(2\phi)} p_F^4 }{4 m^2}
\nonumber\\
& & \hspace{-2.5cm}\times 
\frac{ \left(\frac{1}{2}-\frac{3 a {\cal N}^2}{8\epsilon_F^2} \right)^2 -
4 i s \left(\frac{1}{2}-\frac{3 a {\cal N}^2}{8\epsilon_F^2} \right) }
{\kappa q^2 + i s F_2 N(0)  \sin^2{(2\phi)}},
\end{eqnarray}
where we use the mean field equation,
$F^{-1}_2 - \frac{N(0)}{2} + N(0)\frac{3 a {\cal N}^2}{8 \epsilon_F^2}
= F_2^{-1} - \langle Q_{xx} \rangle/{\cal N} =0$.

The sound attenuation is strongly enhanced via 
its coupling to the Goldstone mode, and it is given by
\begin{equation}
\alpha_l^G \approx \lambda_l  \sin^2{(2\phi)}
\frac{ 4 \kappa q^2 +  \sin^2{(2\phi)}  }
{(\kappa q^2 v_F/C_l)^2  +  F_2^2 N(0)^2  \sin^4{(2\phi)} }
\left(\frac{v_F}{C_l} \right).
\end{equation}
The longitudinal sound wave attenuation is anisotropic, because
the Goldstone mode is the transverse fluctuation of the orientational order. 
The sound wave does not couple to the Goldstone mode
when the wave vector of the sound wave is parallel
to the symmetric directions of the Fermi surface,
i.e., when $\phi = 0$ and $\pi/2$.
The other angle of $\phi$ the sound wave is heavily damped
due to the coupling to the {\it overdamped} Goldstone mode.

The presence of the lattice induces a finite gap in the Goldstone mode, and the
size of the gap depends on the details of the structure of the lattice.
We do not consider a finite gap in the Goldstone mode, which
quantitatively reduces its effect on the damping of the sound attenuation.
However, the main conclusion of the strong and anisotropic damping of the sound
wave does not change, because the effect of Goldstone mode 
in the presence of lattice is similar to that of amplitude mode which
also leads to a large damping of the sound wave compared to that 
of the isotropic Fermi liquid. This is presented in the following subsection.

\subparagraph{coupling to the amplitude mode}
The sound wave couples to the amplitude mode which is 
the fluctuation of the magnitude of the overall Fermi surface distortion.
The amplitude mode also couples to the density, but
this coupling is smaller than the stress tensor, $\tau$
by order of $s^2 = (C_l/v_F^2) << 1 $.
The stress tensor and  the amplitude mode correlation
function is given by
\begin{equation}
\langle [\tau_{ll},\delta {\cal N}_{\parallel} ] \rangle=
\frac{\cos{(2\phi)} N(0) p_F^2}{2m}
 \left[\frac{1}{2}-\frac{9 a {\cal N}^2}{8\epsilon_F^2} -2 i s  +O(s^2) \right],
\end{equation}
and the propagator of the amplitude mode is obtained as\cite{vadim,kim}
\begin{equation}
\langle [\delta {\cal N}_{\parallel}, \delta {\cal N}_{\parallel}] \rangle
=F_2^{-1} -2\Delta F_2^{-1} - i s N(0) \cos^2{(2\phi)},
 +O(s^2),
\end{equation} 
where $2\Delta$ is the gap of the amplitude mode and it is 
given by $  F_2 N(0) \frac{3 a {\cal N}^2}{4 \epsilon_F^2}$.
The coupling to the amplitude mode is  written as
\begin{eqnarray}
\frac{F_2(q) \langle [h_l,\delta {\cal N}_{\parallel}]\rangle
\langle [\delta {\cal N}_{\parallel},h_l]\rangle}{ 1-F_2(q) 
\langle [\delta {\cal N}_{\parallel},\delta {\cal N}_{\parallel}]\rangle}
&=& \frac{F_2 N(0)^2 \cos^2{(2\phi)} p_F^4}{4 m^2}
\nonumber\\
& & \hspace{-2.5cm}\times 
 \frac{ \left(\frac{1}{2}-\frac{9 a {\cal N}^2}{8\epsilon_F^2} \right)^2
 -4 i s \left(\frac{1}{2}-\frac{9 a {\cal N}^2}{8\epsilon_F^2} \right) }
{2 \Delta + \kappa q^2+ i s F_2 N(0)  \cos^2{(2\phi)} }.
\end{eqnarray}
The additional and anisotropic sound attenuation, $\alpha_l^a$, occurs
due to the coupling to the amplitude mode, which is given by
\begin{equation}
\alpha_l^{a} \approx 4 \lambda_l \cos^2{(2\phi)} 
\frac{\left(1 - 2\Delta \right)}{2\Delta + \kappa q^2} 
\left( \frac{C_l}{v_F} \right).
\end{equation}
The extra sound attenuation due to amplitude mode, $\alpha_l^a$ 
depends on the microscopic interaction as well as the presence of
the lattice, but it is clear that it can be large
compared to that of the  isotropic Fermi liquid
which is order of $\lambda_l (C_l/v_F)$.
The coupling of the sound wave to the amplitude mode is maximum,
when the wave vector ${\bf q}$ is along the symmetric directions,
while the coupling of the sound wave to the Goldstone mode is absent 
for these directions.
This is expected because the amplitude mode is the longitudinal
fluctuation of the magnitude of the Fermi surface distortion, 
so its effect on the sound wave is maximum
when the sound wave propagates along the major and minor axes
of the elongated Fermi surface.

\paragraph{Discussion and summary}

The sound wave has been studied to probe a broken symmetry, 
because of its possible coupling to an order parameter
in an ordered state.
In superfluid $^3He$, the extensive theoretical and
experimental studies of the sound wave were carried out
especially because of its applicability of probing the complex 
order parameters.\cite{vollhardt}
The coupling to Anderson-Bogoliubov mode, so-called phason mode,
was studied, and it was found that its effect on the zero sound velocity shift
is order of $(1/s)^2 = (v_F/C_l)^2 << 1$ in $^3He$.\cite{maki}
On the other hand, the phason mode in s-wave BCS superconductors
is pushed up to the plasma frequency
due to the  presence of a long range Coulomb interaction.\cite{anderson}
The coupling to the clapping mode in a p-wave superconductors
with broken time reversal symmetry state was studied,
and it was found that its effect on the sound wave velocity
shift is order of $s^2 = (C_l/v_F)^2 << 1$.\cite{kee}

Here we studied the sound wave in the nematic 
Fermi liquid, where the rotational symmetry is spontaneously broken
due to the  quadrupolar density-density interaction
between electrons.
Without taking into account the coupling to the collective modes,
the sound wave velocity shift is reduced in the nematic ordered state.  
Its reduction is anisotropic, and it is maximum when
the sound propagates along the major and the minor axes of 
the elliptical Fermi surface. 
The sound attenuation is weakly damped
which is the same as that in the isotropic Fermi liquid.
%
The  dramatic change in the sound wave was found via its
coupling to the Goldstone mode in the nematic Fermi liquid.
The anisotropic and strong damping of the sound wave occurs in
the nematic ordered state because of the following reasons.
1. The transverse fluctuation of 
the distorted Fermi surface in the nematic Fermi liquid 
couples to the sound wave.
2. The Goldstone mode is overdamped in the nematic ordered state
due to the anisotropic Landau damping.
Therefore, the sound wave becomes heavily damped via its coupling  to
the overdamped Goldstone mode, except when  sound propagates
along the major and minor axes of the elongated Fermi surface.
While the presence of the lattice leads to a finite gap of the Goldstone
mode, the effect of the Goldstone mode on the sound attenuation
is qualitatively the same ---  anisotropic and strong damping of the sound.
The amplitude mode gives the additional and anisotropic damping
of the sound wave. 


A possible relevance of our finding can be found in the thermal
conductivity measurement reported in \cite{buchner}.
The dramatic suppression of the phononic thermal conductivity
was observed in rare earth and Sr doped La$_2$CuO$_4$,
which correlates with the occurrence of superconductivity.\cite{buchner}
The conventional models based on enhanced phonon-defect scattering on alloying 
or conventional electron-phonon scattering fail to explain these observations.
The interpretation of the pronounced damping of the phonons in
the superconducting state was that
the stripe correlations are dynamic which can be viewed as a nearly nematic order
in the superconducting state, while it is either static or absent in
nonsuperconducting state.
The formation of the dynamic stripe correlation was suggested based
the inelastic neutron scattering measurement on
superconducting La$_{2-x}$Sr$_x$CuO$_4$\cite{mason}. 

In summary, we find that the behavior of the sound wave is dramatically different
in the nematic Fermi liquid from that in  the isotropic Fermi liquid.
If the nematic Fermi liquid exists as a part of the rich phase
diagram of a doped Mott insulator proposed in \cite{emery},
the behavior of the sound wave propagation
should reflect the existence of the nematic order ---
the strong damping of the sound wave is accompanied. 

{\it Acknowledgments}
I deeply appreciate Kazumi Maki, Yong Baek Kim, Steven A. Kivelson,
and V. Oganesyan for useful discussions.
This work was supported  by Canadian Institute for
Advanced Research, Canada Research Chair, and Natural Sciences and Engineering
Research Council of Canada.


\begin{references}
\bibitem{note}In this analysis, we do not take into account
the long-range Coulomb interaction. See \cite{vadim,vadim2}.
\end{references}
\end{document}